\documentclass[twocolumn,showpacs,showkeys]{revtex4}
\usepackage{graphicx}
\usepackage{bm}
\usepackage{color}
\usepackage{amsmath}
\usepackage{natbib}

\begin{document}

\title{Novel soliton in dipolar BEC caused by the quantum fluctuations}

\author{Pavel A. Andreev}
\email{andreevpa@physics.msu.ru}
\affiliation{Faculty of physics, Lomonosov Moscow State University, Moscow, Russian Federation, 119991.}

\date{\today}

\begin{abstract}
Solitons in the extended hydrodynamic model of the dipolar Bose-Einstein condensate with quantum fluctuations are considered.
This model includes
the continuity equation for the scalar field of concentration,
the Euler equation for the vector field of velocity,
the pressure evolution equation for the second rank tensor of pressure,
and
the evolution equation for the third rank tensor.
Large amplitude soliton solution caused by the dipolar part of quantum fluctuations is found.
It appears as the bright soliton.
Hence, it is the area of compression of the number of particles.
Moreover, it exists for the repulsive short-range interaction.

\end{abstract}



\maketitle


Dipolar Bose-Einstein condensates (BECs) demonstrate the formation of quantum droplets
\cite{Kadau Pfau Nature 16}, \cite{Ferrier-Barbut PRL 16}, \cite{Baillie PRA 16},
\cite{Bisset PRA 16}, \cite{Wachtler PRA 16 a1}, \cite{Wachtler PRA 16 a2}, \cite{Blakie pra 16},
\cite{Boudjemaa PRA 20}, \cite{Heinonen PRA 19}, \cite{Malomed Phys D 19}, \cite{Shamriz PRA 20},
\cite{Li PRA 19}, \cite{Aybar PRA 19}, \cite{Examilioti JP B 20}, \cite{Miyakawa PRA 20},
\cite{Bottcher arXiv 20 07}, \cite{Bisset arXiv 20 07}, \cite{Wang arXiv 20 02},
\cite{Edmonds arXiv 20 02}, \cite{Baillie PRA 20}.
Each droplet is the small cloud of atoms,
which can be considered as the soliton-like area of increased concentration.
System of droplets is the highly nonlinear structure explained via the dipolar part of the quantum fluctuations
via the generalized Gross-Pitaevskii (GP) equation including the fourth order nonlinearity.

The traditional GP equation contains one nonlinear term
which is proportional to the third degree of the macroscopic wave function \cite{Dalfovo RMP 99}.
This nonlinearity is caused by the major part of the short-range interaction contribution.
In spite the fact that the BECs is the collection of particles in the quantum state with the lowest energy,
a part of particles can exist in the exited states even at the zero temperature.
It happens via the interaction beyond the mean-field approximation.
This phenomenon is called the BEC depletion caused by the quantum fluctuations.
It is studied both theoretically
\cite{Lee PR 57}, \cite{Pitaevskii PRL 98}, \cite{Braaten PRL 99}, \cite{Astrakharchik PRL 05}
and experimentally \cite{Xu PRL 06}, \cite{Altmeyer PRL 07}, \cite{Papp PRL 08}.
If one considers the quantum fluctuations caused by the short-range interaction within the Bogoliubov-de Gennes theory
one can find additional nonlinear term in the GP equation which is also caused by the short-range interaction,
but it is proportional to the fourth degree of the macroscopic wave function.
It contains same interaction constant as the traditional GP equation.

The dipolar BEC has long history of study over the last 20 years
\cite{Goral PRA 00}, \cite{Santos PRL 00}, \cite{Yi PRA 00},
\cite{Griesmaier PRL 05}, \cite{Wang Nat 10}, \cite{Carr Ye NJP 09}, \cite{Carr DeMille NJP 09},
\cite{Wilson arxiv 11}, \cite{Lahaye Nat 07}, \cite{Lahaye RPP 09}.
The nonsuperfluid fermionic dipolar gases are also considered in literature \cite{Lima PRA 10 a}, \cite{Lima PRA 10 b}.
Traditionally the condensate depletion is studied in terms of Bogoliubov-de Gennes theory
\cite{Lima PRA 11}, \cite{Lima PRA 12}, \cite{Blakie PRA 13}.
It includes the depletion of the dipolar BECs.
However, here we present the microscopic many-particle quantum hydrodynamic theory of the quantum fluctuations.

Although, the dipole-dipole interaction modifies properties of nonlinear structure in BECs
including the width and the amplitude of the bright and dark solitons in the BECs \cite{Andreev EPJ D 14}.
To the best of our knowledge,
the dipole-dipole interaction brings no novel soliton formation.
In Ref. \cite{Andreev EPJ D 14} the analytical solutions are demonstrated for the dipole-dipole interaction of the point-like objects.
Here, it is found that the dipolar part of quantum fluctuations causes the soliton solution.

Recently, the microscopically justified quantum hydrodynamic model describes the quantum fluctuations
via the equations additional to the Euler and continuity equations \cite{Andreev 2005}, \cite{Andreev 2007}.
The interparticle interaction creates the source of the third rank tensor $Q^{\alpha\beta\gamma}$,
which gives the nonzero value of kinetic pressure $p^{\alpha\beta}$.
Both the kinetic pressure and the third rank tensor $Q^{\alpha\beta\gamma}$ are related to the occupation of the excited states.
Their nonzero values for the BECs are the consequence of the quantum fluctuations in BECs.
The pressure evolution equation contains no trace of the interaction.
Hence, its value depends on the third rank tensor $Q^{\alpha\beta\gamma}$ only.
The third rank tensor evolution equation contains the gradient of the concentration square $n^{2}$
multiplied by the additional interaction constant for the short-range interaction.
The long-range dipole-dipole interaction leads to the third derivative of the macroscopic potential of the dipole-dipole interaction.
However, no contribution of the external field, such as the trapping potential, is present in this equation.

The description of collisions of solitons in BECs requires models obtained beyond the mean-field approximation
\cite{Katsimiga NJP 17 01}, \cite{Katsimiga NJP 17 02}, \cite{Katsimiga PRA 18}, \cite{Mistakidis NJP 18}.
In this paper we demonstrate that the described above beyond mean-field model provides a novel soliton solution in dipolar BECs.

Novel soliton solution in dipolar BECs is found in terms of the extended quantum hydrodynamic model,
where the continuity equation for the scalar field of concentration $n$,
the Euler equation for the vector field of velocity $\textbf{v}$,
the pressure evolution equation for the second rank tensor of pressure $p^{\alpha\beta}$,
and
the evolution equation for the third rank tensor $Q^{\alpha\beta\gamma}$ are used.
In this model, the quantum fluctuations are presented in the equation for the third rank tensor evolution.
This model represents the microscopic motion of the quantum particles via the functions describing the collective dynamics
\cite{Koide PRC 13}, \cite{Andreev EPL 16}, \cite{Andreev APL 16}, \cite{Andreev 2001}, \cite{Maksimov QHM 99},
\cite{Andreev PRB 11}, \cite{MaksimovTMP 2001}, \cite{Renziehausen PTEP},
which is described by the many-particle Schrodinger equation:
$$\imath\hbar\partial_{t}\Psi(R,t)=\Biggl[\sum_{i=1}^{N}\biggl(\frac{\hat{\textbf{p}}^{2}_{i}}{2m}+V_{ext}(\textbf{r}_{i},t)\biggr)$$
\begin{equation}\label{BECdqfS20 Hamiltonian micro}
+\frac{1}{2}\sum_{i,j\neq i}U_{ij}+\frac{1}{2}\mu^{2}\sum_{i,j\neq i}\frac{1-3r_{z,ij}^{2}/r_{ij}^{2}}{r_{ij}^{3}} \Biggr]\Psi(R,t),\end{equation}
where $m$ is the mass of the atom,
$\hat{\textbf{p}}_{i}=-\imath\hbar\nabla_{i}$ is the momentum of i-th particle,
$\hbar$ is the Planck constant,
$\mu$ is the magnetic moment of the atom,
$\Psi(R,t)$ is the wave function for the system of $N$ quantum particles,
$R=\{\textbf{r}_{1}, ..., \textbf{r}_{N}\}$,
$V_{ext}$ is the external potential.
The short-range part of boson-boson interaction is presented via potential $U_{ij}=(r_{ij})$,
where $r_{ij}=\mid \textbf{r}_{ij}\mid$,
and $\textbf{r}_{ij}=\textbf{r}_{i}-\textbf{r}_{j}$.
The long-range dipole-dipole interaction of align dipoles
is presented by the last term of the Schrodinger equation (\ref{BECdqfS20 Hamiltonian micro}).

Transition to the description of the collective motion of bosons is made
via the introduction of the concentration (number density)
\cite{Andreev 2001}, \cite{Andreev 1912}, \cite{Andreev PRA08}, \cite{Andreev LP 19}:
\begin{equation}\label{BECdqfS20 concentration def b} n=\int
dR\sum_{i=1}^{N}\delta(\textbf{r}-\textbf{r}_{i})\Psi^{*}(R,t)\Psi(R,t).\end{equation}
The integral in equation (\ref{BECdqfS20 concentration def b}) contains the element of volume in $3N$ dimensional space
$dR=\prod_{i=1}^{N}d\textbf{r}_{i}$.

The derivation \cite{Andreev PRA08} shows that
the concentration (\ref{BECdqfS20 concentration def b}) obeys the continuity equation
\begin{equation}\label{BECdqfS20 cont eq via v} \partial_{t}n+\nabla\cdot (n\textbf{v})=0. \end{equation}

The current of particles is proportional to the momentum density and presented by the following equation
$$\textbf{j}
=\int dR\sum_{i=1}^{N}\delta(\textbf{r}-\textbf{r}_{i})\times$$
\begin{equation}\label{BECdqfS20 j def}
\times\frac{1}{2m_{i}}(\Psi^{*}(R,t)\hat{\textbf{p}}_{i}\Psi(R,t)+c.c.),\end{equation}
with $c.c.$ is the complex conjugation.
The current allows to define the velocity vector field:
$\textbf{v}=\frac{\textbf{j}}{n}$.

The evolution of the current (\ref{BECdqfS20 j def}) follows from the Schrodinger equation (\ref{BECdqfS20 Hamiltonian micro})
and can be presented by the Euler equation:
$$mn\partial_{t}v^{\alpha} +mn(\textbf{v}\cdot\nabla)v^{\alpha}
-\frac{\hbar^{2}}{2m}n\nabla^{\alpha}\frac{\triangle \sqrt{n}}{\sqrt{n}}
+\partial_{\beta}T_{qf}^{\alpha\beta}$$
\begin{equation}\label{BECdqfS20 Euler bosons BEC}
+n\partial^{\alpha}V_{ext}=-g n\partial^{\alpha}n
-n\partial^{\alpha} \Phi_{d},\end{equation}
where $\triangle=\partial_{\beta}\partial_{\beta}$,
and the Einstein's rule for the summation on the repeating subindex is applied.

Major contribution of the short-range interaction appears in the mean-field approximation \cite{Dalfovo RMP 99}
corresponding to the first order by the interaction radius \cite{Andreev PRA08}.
It is presented by the first term on the right-hand side of the Euler equation (\ref{BECdqfS20 Euler bosons BEC}).
It contains the interaction constant $g$ expressed via the potential:
\begin{equation} \label{BECdqfS20 def g} g=\int d\textbf{r}U(r). \end{equation}

The long-range of interaction is presented in the correlationless form corresponding to the main contribution of the interaction
via the macroscopic electrostatic potential:
\begin{equation} \label{BECdqfS20 dd int pot def}  \Phi_{d}=\mu^{2}\int d\textbf{r}'
\frac{1}{|\textbf{r}-\textbf{r}'|^{3}}\biggl(1-3\frac{(z-z')^{2}}{|\textbf{r}-\textbf{r}'|^{2}}\biggr)
n(\textbf{r}',t). \end{equation}

The dipole-dipole interaction presented by potential (\ref{BECdqfS20 dd int pot def}) is not full dipole-dipole interaction,
but the long-range asymptotics of atom-atom interaction.
Potential (\ref{BECdqfS20 dd int pot def}) satisfies the following differential equation
\begin{equation} \label{BECdqfS20 dd int pot def Differential form}  \triangle\Phi_{d}=4\pi\mu^{2}(\partial_{z}^{2}n-\triangle n) , \end{equation}
which is the modification of the Poisson equation.
Equation (\ref{BECdqfS20 dd int pot def Differential form}) is derived
for the arbitrary directions of the pair of dipoles with the further transition to the pair of aligned dipoles.

The evolution of the particles current $\textbf{j}$ (\ref{BECdqfS20 j def}) leads to the flux of momentum $\Pi^{\alpha\beta}$
which is defined as follows
$$\Pi^{\alpha\beta}=\int dR\sum_{i=1}^{N}\delta(\textbf{r}-\textbf{r}_{i}) \frac{1}{4m^{2}}
[\Psi^{*}(R,t)\hat{p}_{i}^{\alpha}\hat{p}_{i}^{\beta}\Psi(R,t)$$
\begin{equation} \label{BECdqfS20 Pi def} +\hat{p}_{i}^{\alpha *}\Psi^{*}(R,t)\hat{p}_{i}^{\beta}\Psi(R,t)+c.c.]. \end{equation}
Equation (\ref{BECdqfS20 Euler bosons BEC}) contains the momentum flux represented via the velocity field.

The continuity equation and the Euler equation are presented via the velocity field.
Transition of the general equations to this form can be made by the representation of the macroscopic wave function
$\Psi(R,t)$ via the real functions $\Psi(R,t)=a(R,t) \exp(\imath S(R,t)/\hbar)$.
The real functions can be called
the amplitude of the wave function $a(R,t)$,
and
the phase of the wave function $S(R,t)$.
The gradient of the wave function gives the velocity of the quantum particle:
$\textbf{v}_{i}(R,t)=\nabla_{i}S(R,t)/m_{i}$.
The deviation of the velocity of quantum particle from the average velocity
$\textbf{u}_{i}(\textbf{r},R,t)=\textbf{v}_{i}(R,t)-\textbf{v}(\textbf{r},t)$
can be called the thermal velocity,
or it is the velocity in the local comoving frame.

Final expression for the momentum flux is obtained in the following form:
\begin{equation}\label{BECdqfS20 Pi via n v p T}
\Pi^{\alpha\beta}=nv^{\alpha}v^{\beta}+p^{\alpha\beta}+T^{\alpha\beta}, \end{equation}
where the tensor function $p^{\alpha\beta}$ in equation (\ref{BECdqfS20 Pi via n v p T}) is the kinetic pressure
\begin{equation}\label{BECdqfS20 def pressure} p^{\alpha\beta}(\textbf{r},t)=\int dR\sum_{i=1}^{N}\delta(\textbf{r}-\textbf{r}_{i})a^{2}(R,t)m_{i}u^{\alpha}_{i}u^{\beta}_{i},\end{equation}
and the simplified form of the second rank tensor $T^{\alpha\beta}$ is found as follows:
\begin{equation} \label{BECdqfS20 Bom2 sigle state}
T^{\alpha\beta}=
-\frac{\hbar^{2}}{4m}\biggl(\partial^{\alpha}\partial^{\beta}n
-\frac{1}{n}(\partial^{\alpha}n)(\partial^{\beta}n)\biggr)
+T^{\alpha\beta}_{qf}.\end{equation}

The BEC is the collection of particles in the quantum state with the lowest energy.
It corresponds to the zero temperature $T=p^{\beta\beta}/3n$.
Therefore, the zero kinetic pressure is the characteristic of the BEC.
However, we can consider the pressure evolution equation.
Its derivation is made for the arbitrary distribution of particles over quantum states.
It corresponds to the arbitrary temperatures.
Transition to the zero temperature is made after derivation of the general equation.

Hence, the derivation of the momentum flux evolution equation is made by the consideration of the time derivative of function
$\Pi^{\alpha\beta}$ (\ref{BECdqfS20 Pi def}).
Next, we introduce the velocity field in accordance with the method shown before equation (\ref{BECdqfS20 Pi via n v p T}).
The final equation reduces to the equation for the part of the quantum Bohm potential $T^{\alpha\beta}$ (\ref{BECdqfS20 Bom2 sigle state})
caused by the quantum fluctuations $T_{qf}^{\alpha\beta}$
(it can be also interpreted as the part of pressure caused by the quantum fluctuations):
\begin{equation} \label{BECdqfS20 eq evolution T qf}
\partial_{t}T_{qf}^{\alpha\beta} +\partial_{\gamma}(v^{\gamma}T_{qf}^{\alpha\beta})
+T_{qf}^{\alpha\gamma}\partial_{\gamma}v^{\beta}
+T_{qf}^{\beta\gamma}\partial_{\gamma}v^{\alpha}
+\partial_{\gamma}Q_{qf}^{\alpha\beta\gamma}=0.  \end{equation}

All terms in equation (\ref{BECdqfS20 eq evolution T qf}) are proportional to $T_{qf}^{\alpha\beta}$
and the flux of pressure in the comoving frame $Q_{qf}^{\alpha\beta\gamma}$.
Hence, tensor $Q_{qf}^{\alpha\beta\gamma}$ goes to zero together with the kinetic pressure at the zero temperature.
Hence, equation (\ref{BECdqfS20 eq evolution T qf}) gives the identity 0=0.
However, it is the quasi-classical description of the pressure evolution equation.
To understand the full quantum picture,
we need to derive the equation for the third rank tensor evolution $Q_{qf}^{\alpha\beta\gamma}$ at the arbitrary temperature.
Afterwords, we make the transition to the zero temperature in the derived equation.
As the result we get equation (\ref{BECdqfS20 eq evolution Q qf}) presented below.
Equation (\ref{BECdqfS20 eq evolution Q qf}) shows that $\partial_{t}Q_{qf}^{\alpha\beta\gamma}\neq 0$ at the zero temperature
due to the nonzero contribution of the interaction.
Hence, there is the interaction related source of $Q_{qf}^{\alpha\beta\gamma}$
and consequently it gives the source of $T_{qf}^{\alpha\beta}$ via the pressure evolution equation.
Therefore, the interaction causes the occupation of the quantum states with nonminimal energies.
This description corresponds to the well-known nature of the quantum fluctuations
\cite{Pitaevskii PRL 98}, \cite{Lima PRA 11}, \cite{Lima PRA 12}, \cite{Blakie PRA 13}.

No interaction gives contribution in the pressure evolution equation (\ref{BECdqfS20 eq evolution T qf}).
The form of the trapping potential does not affect the pressure evolution.

The evolution of the second rank tensor of pressure leads to the flux of the momentum flux:
$$M^{\alpha\beta\gamma}=\int dR\sum_{i=1}^{N}\delta(\textbf{r}-\textbf{r}_{i}) \frac{1}{8m_{i}^{3}}\biggl[\Psi^{*}(R,t)\hat{p}_{i}^{\alpha}\hat{p}_{i}^{\beta}\hat{p}_{i}^{\gamma}\Psi(R,t)$$
$$+\hat{p}_{i}^{\alpha *}\Psi^{*}(R,t)\hat{p}_{i}^{\beta}\hat{p}_{i}^{\gamma}\Psi(R,t)
+\hat{p}_{i}^{\alpha *}\hat{p}_{i}^{\gamma *}\Psi^{*}(R,t)\hat{p}_{i}^{\beta}\Psi(R,t)$$
\begin{equation} \label{BECdqfS20 M alpha beta gamma def}
+\hat{p}_{i}^{\gamma *}\Psi^{*}(R,t)\hat{p}_{i}^{\alpha}\hat{p}_{i}^{\beta}\Psi(R,t)+c.c.\biggr]. \end{equation}

Calculations give the following representation of the third rank tensor $M^{\alpha\beta\gamma}$
via the velocity field and other hydrodynamic functions:
$$M^{\alpha\beta\gamma}= nv^{\alpha}v^{\beta}v^{\gamma} +v^{\alpha} p^{\beta\gamma} +v^{\beta} p^{\alpha\gamma} $$
\begin{equation} \label{BECdqfS20 M via nv pv QTL} +v^{\gamma} p^{\alpha\beta}+Q^{\alpha\beta\gamma}
+T^{\alpha\beta\gamma}, \end{equation}
where we have two new functions.
One of them the quasi-classic third rank tensor in the comoving frame:
\begin{equation}\label{BECdqfS20 def Q} Q^{\alpha\beta\gamma}(\textbf{r},t)=\int dR\sum_{i=1}^{N}\delta(\textbf{r}-\textbf{r}_{i})a^{2}(R,t)u^{\alpha}_{i}u^{\beta}_{i}u^{\gamma}_{i} .\end{equation}

The quantum part of the tensor $M^{\alpha\beta\gamma}$ is found:
$$T^{\alpha\beta\gamma}
=\frac{\hbar^{2}}{2m^{2}}\biggl[-\frac{1}{6}n(\partial^{\alpha}\partial^{\beta} v^{\gamma}
+\partial^{\alpha}\partial^{\gamma} v^{\beta}
+\partial^{\beta}\partial^{\gamma} v^{\alpha})$$
\begin{equation} \label{BECdqfS20 T alpha beta gamma explicit}
+T^{\beta\gamma}\cdot v^{\alpha}
+T^{\alpha\beta}\cdot v^{\gamma}
+T^{\alpha\gamma}\cdot v^{\beta}\biggr] .\end{equation}

Equation for the evolution of quantum-thermal part of the third rank tensor is \cite{Andreev 2005}, \cite{Andreev 2007}:
$$\partial_{t}Q_{qf}^{\alpha\beta\gamma} +\partial_{\delta}(v^{\delta}Q_{qf}^{\alpha\beta\gamma})
+Q_{qf}^{\alpha\gamma\delta}\partial_{\delta}v^{\beta}
+Q_{qf}^{\beta\gamma\delta}\partial_{\delta}v^{\alpha}
+Q_{qf}^{\alpha\beta\delta}\partial_{\delta}v^{\gamma}$$
$$=\frac{\hbar^{2}}{4m^{2}} n\biggl(g_{2}I_{0}^{\alpha\beta\gamma\delta}\partial^{\delta}n
+\partial^{\alpha}\partial^{\beta}\partial^{\gamma}\Phi_{d}\biggr)$$
\begin{equation} \label{BECdqfS20 eq evolution Q qf}
+\frac{1}{mn}(T_{qf}^{\alpha\beta}\partial^{\delta}T_{qf}^{\gamma\delta}
+T_{qf}^{\alpha\gamma}\partial^{\delta}T_{qf}^{\beta\delta}
+T_{qf}^{\beta\gamma}\partial^{\delta}T_{qf}^{\alpha\delta}),  \end{equation}
where
\begin{equation} \label{BECdqfS20 I 4}
I_{0}^{\alpha\beta\gamma\delta}=\delta^{\alpha\beta}\delta^{\gamma\delta} +\delta^{\alpha\gamma}\delta^{\beta\delta}+\delta^{\alpha\delta}\delta^{\beta\gamma}. \end{equation}

The main contribution of the short-range interaction is proportional to the second interaction constant:
\begin{equation} \label{BECdqfS20 def g 2} g_{2}=\frac{2}{3}\int d\textbf{r} U''(r). \end{equation}
It is obtained in the first order by the interaction radius \cite{Andreev PRA08}.

Different versions of the extended hydrodynamics for various physical systems are presented in Refs.
\cite{Andreev 2001}, \cite{Andreev 1912}, \cite{Tokatly PRB 99}, \cite{Tokatly PRB 00}, \cite{Andreev 2003}.
Novel approaches to the development of hydrodynamics are recently presented in Refs. \cite{Bertini PRL 16} and \cite{Ruggiero QGH arxiv 19}.

Consider solitons in the uniform boundless BEC with no restriction on the amplitude of soliton.
Hence, we have no external potential $V_{ext}=0$.

\textit{Consider the simplified form of the hydrodynamic equations giving main contribution in the soliton solution.}
The continuity equation (\ref{BECdqfS20 cont eq via v}) requires no simplification.
No simplification of the equation of field (\ref{BECdqfS20 dd int pot def Differential form}) is required too.

We drop the traditional part of the quantum Bohm potential in the Euler equation (\ref{BECdqfS20 Euler bosons BEC}):
$$mn\partial_{t}v^{\alpha} +mn(\textbf{v}\cdot\nabla)v^{\alpha}
+\partial_{\beta}T_{qf}^{\alpha\beta}$$
\begin{equation}\label{BECdqfS20 Euler bosons BEC simple}
=-g n\partial^{\alpha}n
-n\partial^{\alpha} \Phi_{d}.\end{equation}

The pressure evolution equation simplifies to two terms:
\begin{equation} \label{BECdqfS20 eq evolution T qf simple}
\partial_{t}T_{qf}^{\alpha\beta} +\partial_{\gamma}Q_{qf}^{\alpha\beta\gamma}=0,  \end{equation}
where
the quantum Bohm potential $T_{qf}^{\alpha\beta}$ is caused purely by the flux $Q_{qf}^{\alpha\beta\gamma}$.

Equation for the evolution of quantum-thermal part of the third rank tensor is:
\begin{equation} \label{BECdqfS20 eq evolution Q qf simple}
\partial_{t}Q_{qf}^{\alpha\beta\gamma}
=\frac{\hbar^{2}}{4m^{2}} n\biggl(g_{2}I_{0}^{\alpha\beta\gamma\delta}\partial^{\delta}n
+\partial^{\alpha}\partial^{\beta}\partial^{\gamma}\Phi_{d}\biggr),  \end{equation}
where
the interaction on the right-hand sides represents the quantum fluctuations.
It is assumed that
the evolution of the tensor $Q_{qf}^{\alpha\beta\gamma}$ is mainly caused by the interaction.

There is no independent source of interaction in the pressure evolution equation (\ref{BECdqfS20 eq evolution Q qf simple}).
Hence, the third rank tensor $Q_{qf}^{\alpha\beta\gamma}$ is the single source of interaction in the pressure evolution equation.
As it is mentioned above, we focus on the pressure or the quantum Bohm potential
caused by the quantum fluctuations.
Hence, we consider the interaction caused $T_{qf}^{\alpha\beta}$ and drop the kinematic terms
(see  equation (\ref{BECdqfS20 eq evolution Q qf simple})).


We consider the one dimensional solution.
We chose the direction of wave propagation perpendicular to the direction of titled dipoles.
We seek the stationary solutions of the nonlinear equations.
We assume the steady state in the comoving frame.
Therefore, the dependence of the time and space coordinates is combined in the single variable
$\xi=x-u t$.
Parameter $u$ is the constant velocity of the nonlinear solution.
Therefore, all hydrodynamic functions depend on $\xi$ and $u$.
We also assume that perturbations vanish at $\xi\rightarrow\pm\infty$.
It gives the following reduction of equations (\ref{BECdqfS20 cont eq via v}), (\ref{BECdqfS20 dd int pot def Differential form}),
(\ref{BECdqfS20 Euler bosons BEC simple}), (\ref{BECdqfS20 eq evolution T qf simple}), (\ref{BECdqfS20 eq evolution Q qf simple}).

Equation (\ref{BECdqfS20 eq for second derivative n or phi}) can be integrated to obtain the "energy integral" in the following manner
\begin{equation}\label{BECdqfS20 final eq for n or phi OT}
\frac{1}{2}(\partial_{\xi}n)^{2} +\frac{m^{2}u^{2}}{\pi\mu^{2}\hbar^{2}}V_{eff}(n)=0,\end{equation}
where $V_{eff}(n)$ is the Sagdeev potential
\cite{Sagdeev RPP 66}, \cite{Schamel PF 77}, \cite{Witt PF 83}, \cite{Mamun PRE 97}, \cite{Shah PoP 10},
\cite{Marklund PRE 07}, \cite{Akbari-Moghanjoughi PP 17}:
\begin{equation}\label{BECdqfS20 Sagdeev potential OT}
V_{eff}(n)=\frac{1}{2}\biggl[
\biggl(g-4\pi\mu^{2}-\frac{1}{u^{2}}\frac{\hbar^{2}}{4m^{2}} 3g_{2}-\frac{mu^{2}}{n}\biggr)\biggr](n-n_{0})^{2}
.\end{equation}
Details of derivation of equations (\ref{BECdqfS20 final eq for n or phi OT}) and (\ref{BECdqfS20 Sagdeev potential OT})
are presented in the Supplementary Materials.
Moreover, the estimation of the area of applicability of equations
(\ref{BECdqfS20 Euler bosons BEC simple}), (\ref{BECdqfS20 eq evolution T qf simple}), and (\ref{BECdqfS20 eq evolution Q qf simple})
is discussed in Supplementary Materials either.

In order the soliton to exist, the effective potential $V_{eff}(n)$ should have a local maximum in the point $n=0$.
Moreover, equation $V_{eff}(n)=0$ should have at least one real solution $n'\neq0$.
This value of concentration $n'$ determines the amplitude $n'$ of the soliton as the function of velocity $u$.

Equation $V_{eff}=0$ can be solved analytically for $n\neq n_{0}$:
\begin{equation}\label{BECdqfS20 Soliton Amplitude}
n'=\frac{mu^{2}}{g-4\pi\mu^{2}-\frac{1}{u^{2}}\frac{\hbar^{2}}{4m^{2}} 3g_{2}}.\end{equation}

Equation (\ref{BECdqfS20 Sagdeev potential}) allows to introduce the effective dimensionless interaction constant
$G=(n_{0}/mu^{2})(g-4\pi\mu^{2}-3\hbar^{2}g_{2}/4m^{2}u^{2})$.

Let us consider the limit of the small dipole-dipole interaction and the small short-range part of the quantum fluctuations
in compare with the mean-field of the short-range interaction presented by the Gross-Pitaevskii interaction constant $g$.
Hence, we have $G\approx2g/mu^{2}$.
We have soliton solution for the positive interaction constant $G\sim g>0$
which corresponds to the repulsive short-range interaction.

The solitary waves exist due to the balance between the nonlinearities
caused by the GP interaction, dipole-dipole interaction, and short-range part of quantum fluctuations
and the dispersion
induced by the dipolar part of quantum fluctuations.
The dipole-dipole interaction and short-range part of quantum fluctuations introduce an
additional negative contribution to the interaction constant of the GP approximation.

To understand the possibility of the soliton existence in the regime under consideration
we study the form of the Sagdeev potential (\ref{BECdqfS20 Sagdeev potential}).

\begin{figure}
\includegraphics[width=8cm,angle=0]{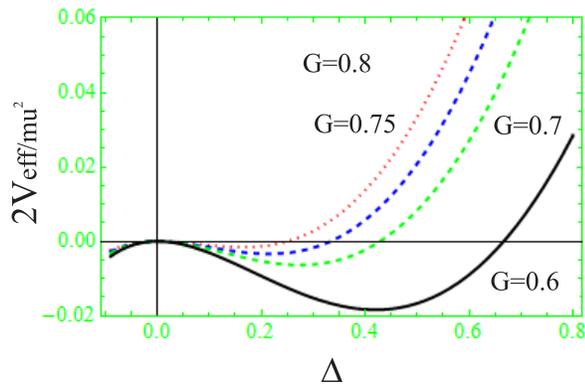}
\caption{\label{BECdqfS20 Fig 01}
The Sagdeev potential as the function of the dimensionless deviation $\Delta=(n-n_{0})/n_{0}$ of the concentration $n$ from
the equilibrium value $n_{0}$ (the concentration at the infinity $n_{0}=n(x=\pm\infty)$)
is demonstrated for the value of $\Delta$ below 1 in accordance with the area of applicability of the simplified hydrodynamic equations in accordance with Supplementary materials.
The Sagdeev potential depends on one parameter $G$,
which is the combination of the interaction constants including the effective interaction constant for the dipole-dipole interaction $g_{d}=4\pi\mu^{2}$.
The figure shows that
there is the "potential gap" in the area of positive deviations $\Delta$.
It means that there is a bright soliton (the area of increased concentration).}
\end{figure}

The Sagdeev potential (\ref{BECdqfS20 Sagdeev potential}) is plotted in Fig. (\ref{BECdqfS20 Fig 01})
for the positive effective interaction constant.
This regime is chosen since it shows the existence of the soliton solution.

\begin{figure}
\includegraphics[width=8cm,angle=0]{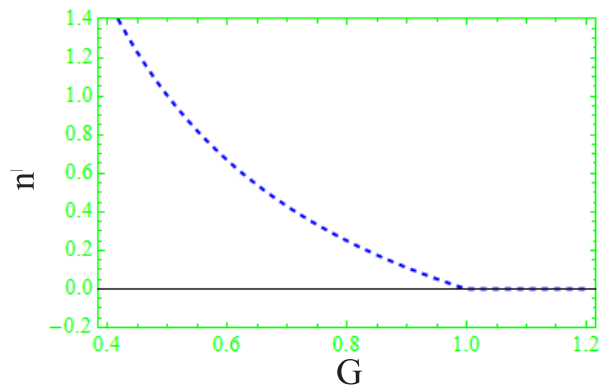}
\caption{\label{BECdqfS20 Fig 02}
The dimensionless soliton amplitude value for concentration $n'/n_{0}$ as the function of the effective interaction constant is demonstrated.
This dependence is given analytically by equation (\ref{BECdqfS20 Soliton Amplitude}).
Here, the decrease of amplitude $n'$ corresponds to the decrease of area of negative potential in Fig. \ref{BECdqfS20 Fig 01}.
Decrease of $G$ gives the monotonic increase of amplitude $n'$.
Hence, the area of applicability of obtained solution can be broken.
Therefore, this figure shows the restrictions on the effective interaction constant $G$.}
\end{figure}

Figures (\ref{BECdqfS20 Fig 01}) and (\ref{BECdqfS20 Fig 02}) show that
the amplitude of soliton (the value $n'$, where $V(n')=0$) decreases with the increase of the effective interaction constant $G$.
The increase of the velocity of soliton $u$ decreases constant $G$ and, consequently, decreases the amplitude.
The increase of the soliton velocity $u$ diminish the role of the short-range interaction part of the quantum fluctuations
in the effective interaction constant.

The traditional bright and dark solitons in neutral atomic BEC are caused by the nonlinearity created
by the short-range interaction in the Gross-Pitaevskii approximation.
There are different generalizations of the Gross-Pitaevkii model include effects beyond the mean-field approximation.
An example of the beyond mean-field model of BEC has been derived in this paper.
The quantum fluctuations have been included here via the extended hydrodynamic model
which includes
the continuity equation for the scalar field of concentration,
the Euler equation for the vector field of velocity,
the pressure evolution equation for the second rank tensor of pressure,
and
the evolution equation for the third rank tensor.

The dipolar part of quantum fluctuations is presented by the term proportional to the third derivative of the electrostatic potential.
For one dimensional perturbations in the single fluid species the potential is proportional to the variation of the concentration from the equilibrium value.

Hence, the high derivatives of the concentration appears in term presenting the dipolar part of quantum fluctuations.


We acknowledge that the work is supported by the Russian Foundation for Basic Research (grant no. 20-02-00476).

\newpage

\section{Supplementary materials}

\subsection{One dimensional limit of hydrodynamic equations}

We consider the one dimensional solution.
We chose the direction of wave propagation perpendicular to the direction of titled dipoles.
We seek stationary solutions of the nonlinear equations.
We assume the steady state in the comoving frame.
Therefore, the dependence of the time and space coordinates is combined in the single variable
$\xi=x-u t$.
Parameter $u$ is the constant velocity of the nonlinear solution.
Therefore, all hydrodynamic functions depend on $\xi$ and $u$.
We also assume that perturbations vanish at $\xi\rightarrow\pm\infty$.
It gives the following reduction of equations (\ref{BECdqfS20 cont eq via v}), (\ref{BECdqfS20 dd int pot def Differential form}),
(\ref{BECdqfS20 Euler bosons BEC simple}), (\ref{BECdqfS20 eq evolution T qf simple}), (\ref{BECdqfS20 eq evolution Q qf simple}).

The reduced continuity equation is
\begin{equation}\label{BECdqfS20 cont simple 1D} -u\partial_{\xi}n+\partial_{\xi}(nv^{x})=0, \end{equation}
where
the time derivatives $\partial_{t}$ is replaced by $-u\partial_{\xi}$ in accordance with the variable $\xi$ introduced for the stationary solution.

The reduced Euler equation is
$$-umn\partial_{\xi}v^{x} +mnv^{x}\partial_{\xi}v^{x}
+\partial_{\xi}T_{qf}^{xx}$$
\begin{equation}\label{BECdqfS20 Euler bosons BEC simple 1D}
=-g n\partial_{\xi}n
-n\partial_{\xi} \Phi_{d}.\end{equation}

The Poisson equation (\ref{BECdqfS20 dd int pot def Differential form}) transforms to
\begin{equation} \label{BECdqfS20 dd int pot def simple 1D}
\partial_{\xi}^{2}\Phi_{d}=-4\pi\mu^{2}\partial_{\xi}^{2} n . \end{equation}

The reduced pressure evolution equation is
\begin{equation} \label{BECdqfS20 eq evolution T qf simple 1D}
-u\partial_{\xi}T_{qf}^{xx} +\partial_{\xi}Q_{qf}^{xxx}=0. \end{equation}

Equation for the evolution of quantum-thermal part of the third rank tensor is:

\begin{equation} \label{BECdqfS20 eq evolution Q qf simple 1D}
-u\partial_{\xi}Q_{qf}^{xxx}
=\frac{\hbar^{2}}{4m^{2}} n\biggl(3g_{2}\partial_{\xi}n
+\partial_{\xi}^{3}\Phi_{d}\biggr), \end{equation}
where
we have the source for the dispersion of the nonlinear wave presented by the third derivative of the potential $\Phi_{d}$ (\ref{BECdqfS20 dd int pot def}).

The neglecting of the quantum Bohm potential (\ref{BECdqfS20 Bom2 sigle state})
in the Euler equation corresponds to the following restriction for the velocities of the soliton propagation $u^{2}\ll\mu^{2}n_{0}$.

\subsection{Derivation of the soliton solution}

The continuity equation (\ref{BECdqfS20 cont simple 1D}) can be integrated
\begin{equation}\label{BECdqfS20 cont simple 1D integrated} n(v^{x}-u)=-un_{0}, \end{equation}
where we use the boundary conditions with nonzero concentration at infinity $n(\xi\rightarrow\pm\infty)=n_{0}$,
and zero value of velocity at infinity $v^{x}(\xi\rightarrow\pm\infty)=0$.

The Euler equation (\ref{BECdqfS20 Euler bosons BEC simple 1D}) can not be integrated at this stage due to the third term,
which is caused by the quantum fluctuations,
which is proportional to $\frac{\partial_{\xi}T_{qf}^{xx}}{n}$.
It will be considered later after the analysis of solution for $T_{qf}^{xx}$.

The derivative of the second rank tensor $T_{qf}^{xx}$ is presented
via the derivative of the third rank tensor $Q_{qf}^{xxx}$ by equation (\ref{BECdqfS20 eq evolution T qf simple 1D}).
Equation (\ref{BECdqfS20 eq evolution T qf simple 1D}) can be integrated,
but it is not necessary
since we need the derivative of the second rank tensor $\partial_{\xi}T_{qf}^{xx}$ for the Euler equation (\ref{BECdqfS20 Euler bosons BEC simple 1D}).

The derivative of the third rank tensor $\partial_{\xi}Q_{qf}^{xxx}$ is found from equation (\ref{BECdqfS20 eq evolution Q qf simple 1D}).
The extra multiplier $n$ in front of the third derivative of the potential $\Phi_{d}$ is canceled in expression $\frac{\partial_{\xi}T_{qf}^{xx}}{n}$,
so the final form of the Euler equation can be integrated.

Equation (\ref{BECdqfS20 dd int pot def simple 1D}) can be integrated twice.
After first integration including the zero derivatives of the concentration and the potential of electric field at infinity
we have
\begin{equation} \label{BECdqfS20 dd int pot def simple 1D int 1}
\partial_{\xi}\Phi_{d}=-4\pi\mu^{2}\partial_{\xi} n . \end{equation}
Next, we make the second integration to find the expression of the potential:
\begin{equation} \label{BECdqfS20 dd int pot def simple 1D int 2}
\Phi_{d}=-4\pi\mu^{2} (n-n_{0}) . \end{equation}

The final expression for the derivative of the second rank tensor is following:
\begin{equation} \label{BECdqfS20 derivative of T qf}
\partial_{\xi}T_{qf}^{xx} =-\frac{1}{u^{2}}\frac{\hbar^{2}}{4m^{2}} n\biggl(3g_{2}\partial_{\xi}n
+\partial_{\xi}^{3}\Phi_{d}\biggr). \end{equation}
It gives the following form of the Euler equation:
$$m(-u)\partial_{\xi}v^{x}
-\frac{1}{u^{2}}\frac{\hbar^{2}}{4m^{2}} \biggl(3g_{2}\partial_{\xi}n
-4\pi\mu^{2}\partial_{\xi}^{3} n\biggr)$$
\begin{equation}\label{BECdqfS20 Euler bosons BEC simple 1D final}
+mv^{x}\partial_{\xi}v^{x}
=-g \partial_{\xi}n
+4\pi\mu^{2}\partial_{\xi} n,\end{equation}
where
the potential of electric field $\Phi_{d}$ is expressed via the concentration in accordance with
equation (\ref{BECdqfS20 dd int pot def simple 1D int 1}).

\begin{figure}
\includegraphics[width=8cm,angle=0]{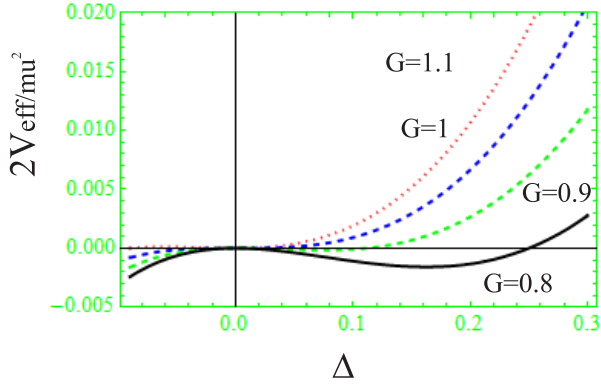}
\caption{\label{BECdqfS20 Fig 03}
The Sagdeev potential as the function of the dimensionless deviation $\Delta=(n-n_{0})/n_{0}$ of the concentration $n$ from
the equilibrium value $n_{0}$ is demonstrated for the relatively large values of the effective interaction constant $G$.
Disappearance of the area of negative potential is demonstrated at the increase of the parameter $G$.}
\end{figure}

We integrate Euler equation (\ref{BECdqfS20 Euler bosons BEC simple 1D final})
$$m(-u)v_{x} +\frac{1}{2}mv_{x}^{2}
-\frac{1}{u^{2}}\frac{\hbar^{2}}{4m^{2}} \biggl(3g_{2}(n-n_{0})
-4\pi\mu^{2}\partial_{\xi}^{2}n \biggr)$$
\begin{equation}\label{BECdqfS20 Euler bosons BEC simple 1D final integrated}
+(g-4\pi\mu^{2}) (n-n_{0})=0 .\end{equation}

The third and fourth terms can be combined together
since they are proportional to $n-n_{0}$.
Hence, the quantum fluctuations caused by the short-range interaction gives the variation of the Gross-Pitaevskii interaction constant $g$.

The dipolar part of the quantum fluctuations plays the crucial contribution in the formation of novel soliton.
Equation (\ref{BECdqfS20 Euler bosons BEC simple 1D final integrated}) becomes the differential equation due to the dipolar part of the quantum fluctuations.
If we drop the dipolar part of the quantum fluctuations
we obtain the constant value of concentration.
It means that no soliton exist in this limit.

Using the integral of the continuity equation (\ref{BECdqfS20 cont simple 1D integrated})
we express the velocity via the concentration.
Hence, equation (\ref{BECdqfS20 Euler bosons BEC simple 1D final integrated}) becomes the equation relatively one function:
\begin{equation}\label{BECdqfS20 eq for second derivative n or phi}\partial_{\xi}^{2}n +\frac{m^{3}u^{2}}{\pi\mu^{2}\hbar^{2}}f(n)=0,\end{equation}
where
$$f(n)=-\frac{1}{2}mu^{2}\frac{n^{2}-n_{0}^{2}}{n^{2}}
-\frac{1}{u^{2}}\frac{\hbar^{2}}{4m^{2}} 3g_{2}(n-n_{0}) $$
\begin{equation}\label{BECdqfS20 pre function for Sagdeev potential}
+(g-4\pi\mu^{2}) (n-n_{0}).\end{equation}

Equation (\ref{BECdqfS20 eq for second derivative n or phi}) can be integrated to obtain the "energy integral" in the following manner
\begin{equation}\label{BECdqfS20 final eq for n or phi} \frac{1}{2}(\partial_{\xi}n)^{2} +\frac{m^{2}u^{2}}{\pi\mu^{2}\hbar^{2}}V_{eff}(n)=0,\end{equation}
where $V_{eff}(n)$ is the Sagdeev potential
$$ V_{eff}(n)=-\frac{1}{2}mu^{2}\frac{(n-n_{0})^{2}}{n}$$
\begin{equation}\label{BECdqfS20 Sagdeev potential}
+\frac{1}{2}\biggl(g-4\pi\mu^{2}-\frac{1}{u^{2}}\frac{\hbar^{2}}{4m^{2}} 3g_{2}\biggr)(n-n_{0})^{2}
.\end{equation}

\begin{figure}
\includegraphics[width=8cm,angle=0]{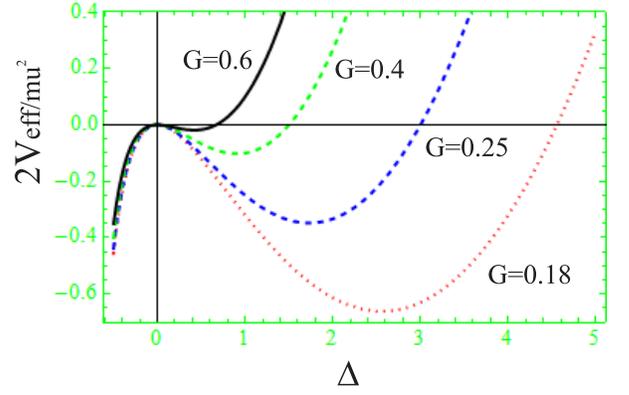}
\caption{\label{BECdqfS20 Fig 04}
The Sagdeev potential as the function of the dimensionless deviation $\Delta=(n-n_{0})/n_{0}$ of the concentration $n$ from
the equilibrium value $n_{0}$ is demonstrated for the relatively large values of $\Delta$.
Here, the amplitude of soliton becomes large enough to brake regime of existence of the simplified hydrodynamic equations.}
\end{figure}

\subsection{Justification of the simplified equations}

Consider full set of equations (\ref{BECdqfS20 cont eq via v}), (\ref{BECdqfS20 Euler bosons BEC}),
(\ref{BECdqfS20 eq evolution T qf}), (\ref{BECdqfS20 eq evolution Q qf})
for the one dimensional case to estimate the contribution of the dropped terms:
\begin{equation}\label{BECdqfS20 cont eq via v 1D nonS}
\partial_{t}n+\partial_{x} (nv^{x})=0, \end{equation}

$$mn\partial_{t}v^{x} +mnv^{x}\partial_{x}v^{x}
+g n\partial_{x}n
+n\partial_{x} \Phi_{d}$$
\begin{equation}\label{BECdqfS20 Euler bosons BEC 1D nonS}
+\partial_{x}T_{qf}^{xx}
=\frac{\hbar^{2}}{2m}n\partial_{x}\frac{\partial_{x}^{2} \sqrt{n}}{\sqrt{n}},\end{equation}
\begin{equation} \label{BECdqfS20 eq evolution T qf 1D nonS}
\partial_{t}T_{qf}^{xx} +\partial_{x}Q_{qf}^{xxx}=-v^{x}\partial_{x}T_{qf}^{xx}
-3T_{qf}^{xx}\partial_{x}v^{x},  \end{equation}
and
$$\partial_{t}Q_{qf}^{xxx} -\frac{\hbar^{2}}{4m^{2}} n\biggl(3g_{2}\partial_{x}n
+\partial_{x}^{3}\Phi_{d}\biggr)$$
\begin{equation} \label{BECdqfS20 eq evolution Q qf 1D nonS}
=-v^{x}\partial_{x}Q_{qf}^{xxx}
-4Q_{qf}^{xxx}\partial_{x}v^{x}
+\frac{3}{mn}T_{qf}^{xx}\partial_{x}T_{qf}^{xx},  \end{equation}
where
the left-hand side in equations (\ref{BECdqfS20 Euler bosons BEC 1D nonS}),
(\ref{BECdqfS20 eq evolution T qf 1D nonS}), (\ref{BECdqfS20 eq evolution Q qf 1D nonS})
contains the terms used in the simplified regime.
The right-hand side gives the terms dropped earlier for their further estimation.

Equation (\ref{BECdqfS20 final eq for n or phi}) shows existence of the soliton solution for the concentration.
Equations (\ref{BECdqfS20 cont simple 1D})-(\ref{BECdqfS20 derivative of T qf}) allows to express other hydrodynamic functions
via the concentration:
\begin{equation} \label{BECdqfS20 v via n}
\tilde{v}_{x}=u\biggl(1-\frac{n_{0}}{n}\biggr).  \end{equation}

Transition to the frame comoving with the soliton
we change $\partial_{t}$ on $-u\partial_{\xi}$ and $\partial_{x}=\partial_{\xi}$.
Equation (\ref{BECdqfS20 eq evolution Q qf simple 1D}) can be integrated
and we obtain corresponding simplified expression of tensor $Q^{xxx}$:
\begin{equation} \label{BECdqfS20 Q solution from simplified eq}
u\tilde{T}_{qf}^{xx}=\tilde{Q}_{qf}^{xxx}=\frac{3\hbar^{2}}{8m^{2}u}g_{2}n^{2}
+\frac{\pi\hbar^{2}\mu^{2}}{m^{2}u}n\partial_{\xi}^{2}n
-\frac{\pi\hbar^{2}\mu^{2}}{2m^{2}u}(\partial_{\xi}n)^{2}, \end{equation}
where
equation (\ref{BECdqfS20 eq evolution T qf simple 1D}) is used to get solution for $T_{qf}^{xx}$ in the simplifies regime.
Next, we use this solution to integrate the right-hand side of equation (\ref{BECdqfS20 eq evolution Q qf 1D nonS}).

The right-hand side of equation (\ref{BECdqfS20 eq evolution Q qf 1D nonS}) contains the following combination of the hydrodynamic functions
$\tilde{v}^{x}\partial_{x}\tilde{Q}_{qf}^{xxx}
+4\tilde{Q}_{qf}^{xxx}\partial_{x}\tilde{v}^{x}$.
Using solution (\ref{BECdqfS20 Q solution from simplified eq}) this combination can be found as the function of the concentration
$$\tilde{v}^{x}\partial_{x}\tilde{Q}_{qf}^{xxx}
+4\tilde{Q}_{qf}^{xxx}\partial_{x}\tilde{v}^{x}=
-\frac{\hbar^{2}}{4m^{2}}(n-n_{0})[3g_{2}\partial_{\xi}n-4\pi\mu^{2}\partial_{\xi}^{3}n]$$
\begin{equation} \label{BECdqfS20 }
-\frac{3\hbar^{2}}{m^{2}}g_{2}n_{0}\frac{(\partial_{\xi}n)^{2}}{n}
+\frac{4\pi\mu^{2}\hbar^{2}n_{0}}{m^{2}}\frac{(\partial_{\xi}n)}{n}\partial_{\xi}^{2}n
-\frac{2\pi\mu^{2}\hbar^{2}n_{0}}{m^{2}}\frac{(\partial_{\xi}n)^{3}}{n^{2}}.\end{equation}

Here, we obtain the generalized expression for the derivative of the third rank tensor
via the concentration
$$\partial_{\xi}Q_{qf}^{xxx} =
-\frac{1}{u}\Biggl\{\frac{\hbar^{2}}{4m^{2}} n\biggl(3g_{2}\partial_{\xi}n
-4\pi\mu^{2}\partial_{\xi}^{3}n\biggr)$$
$$+\frac{\hbar^{2}}{4m^{2}}(n-n_{0})[3g_{2}\partial_{\xi}n-4\pi\mu^{2}\partial_{\xi}^{3}n]
+\frac{3\hbar^{2}}{m^{2}}g_{2}n_{0}(\partial_{\xi}n)$$
\begin{equation} \label{BECdqfS20 der on xi of Q}
-\frac{4\pi\mu^{2}\hbar^{2}n_{0}}{m^{2}}\frac{(\partial_{\xi}n)}{n}\partial_{\xi}^{2}n
+\frac{2\pi\mu^{2}\hbar^{2}n_{0}}{m^{2}}\frac{(\partial_{\xi}n)^{3}}{n^{2}}\Biggr\},  \end{equation}
where
the last term of equation (\ref{BECdqfS20 eq evolution Q qf 1D nonS})
$\frac{3}{mn}T_{qf,0}^{xx}\partial_{x}T_{qf,0}^{xx}$
is dropped since it is proportional to $\hbar^{4}$
(see equation (\ref{BECdqfS20 Q solution from simplified eq})).

Next, we use equation (\ref{BECdqfS20 der on xi of Q}) to find the right-hand side of equation (\ref{BECdqfS20 eq evolution Q qf 1D nonS})
and we use (\ref{BECdqfS20 Q solution from simplified eq}) to find the right-hand side of equation (\ref{BECdqfS20 eq evolution Q qf 1D nonS})
$$\partial_{\xi}T_{qf}^{xx}
=-\frac{1}{u^{2}}\Biggl\{\frac{\hbar^{2}}{4m^{2}} n\biggl(3g_{2}\partial_{\xi}n
-4\pi\mu^{2}\partial_{\xi}^{3}n\biggr)$$
$$+\frac{\hbar^{2}}{2m^{2}}(n-n_{0})[3g_{2}\partial_{\xi}n-4\pi\mu^{2}\partial_{\xi}^{3}n]
+\frac{15\hbar^{2}}{8m^{2}}g_{2}n_{0}(\partial_{\xi}n)$$
\begin{equation} \label{BECdqfS20 partial xi T}
-\frac{7\pi\mu^{2}\hbar^{2}n_{0}}{m^{2}}\frac{(\partial_{\xi}n)}{n}\partial_{\xi}^{2}n
+\frac{7\pi\mu^{2}\hbar^{2}n_{0}}{2m^{2}}\frac{(\partial_{\xi}n)^{3}}{n^{2}}\Biggr\},  \end{equation}
where
the right-hand side is given by
$$\tilde{v}^{x}\partial_{x}\tilde{T}_{qf}^{xx}
+3\tilde{T}_{qf}^{xx}\partial_{x}\tilde{v}^{x}$$
$$=\biggl(1-\frac{n_{0}}{n}\biggr)\partial_{x}\tilde{Q}_{qf}^{xxx}
+3n_{0}\tilde{Q}_{qf}^{xxx}\frac{\partial_{x}n}{n^{2}}$$
$$=-\frac{\hbar^{2}}{4m^{2}u}(n-n_{0})[3g_{2}\partial_{\xi}n-4\pi\mu^{2}\partial_{\xi}^{3}n]$$
\begin{equation} \label{BECdqfS20 } +3n_{0}\frac{\partial_{\xi}n}{n^{2}}
\biggl[\frac{3\hbar^{2}g_{2}}{8m^{2}u}n^{2}
+\frac{\pi\hbar^{2}\mu^{2}}{m^{2}u}n\partial_{\xi}^{2}n
-\frac{\pi\hbar^{2}\mu^{2}}{2m^{2}u}(\partial_{\xi}n)^{2}\biggr]. \end{equation}

Expressions (\ref{BECdqfS20 v via n}) and (\ref{BECdqfS20 partial xi T}) are substituted in the Euler equation (\ref{BECdqfS20 Euler bosons BEC 1D nonS})
to get equation for the concentration $n$.
It gives the generalization of equation (\ref{BECdqfS20 Euler bosons BEC simple 1D final})
$$-umn\partial_{\xi}v^{x} +mnv^{x}\partial_{\xi}v^{x}
+g n\partial_{\xi}n
+n\partial_{\xi} \Phi_{d}$$
$$=\frac{\hbar^{2}}{2m}n\partial_{\xi}\frac{\partial_{\xi}^{2} \sqrt{n}}{\sqrt{n}}
+\frac{1}{u^{2}}\Biggl\{\frac{\hbar^{2}}{4m^{2}} n\biggl(3g_{2}\partial_{\xi}n
-4\pi\mu^{2}\partial_{\xi}^{3}n\biggr)$$
$$+\frac{\hbar^{2}}{2m^{2}}(n-n_{0})[3g_{2}\partial_{\xi}n-4\pi\mu^{2}\partial_{\xi}^{3}n]
+\frac{15\hbar^{2}}{8m^{2}}g_{2}n_{0}(\partial_{\xi}n)$$
\begin{equation}\label{BECdqfS20 Euler bosons BEC with partial xi T}
-\frac{7\pi\mu^{2}\hbar^{2}n_{0}}{m^{2}}\frac{(\partial_{\xi}n)}{n}\partial_{\xi}^{2}n
+\frac{7\pi\mu^{2}\hbar^{2}n_{0}}{2m^{2}}\frac{(\partial_{\xi}n)^{3}}{n^{2}}\Biggr\}
.\end{equation}

The second term on the right-hand side causes the soliton obtained in this paper.
The third term has similar structure, but $(n-n_{0})$ is placed instead of $n$.
Therefore, the third term can be dropped if $n\approx n_{0}$.
Consequently, parameter $\Delta\ll1$ is small.


\begin{thebibliography}{17}








\bibitem{Kadau Pfau Nature 16} H. Kadau, M. Schmitt,	M. Wenzel, C. Wink, T. Maier, I. Ferrier-Barbut,
T. Pfau, Nature \textbf{530}, 194 (2016).

\bibitem{Ferrier-Barbut PRL 16} I. Ferrier-Barbut, H. Kadau, M. Schmitt, M. Wenzel,
and T. Pfau, Phys. Rev. Lett. \textbf{116}, 215301 (2016).


\bibitem{Baillie PRA 16} D. Baillie, R. M. Wilson, R. N. Bisset, and P. B. Blakie, Phys. Rev. A \textbf{94}, 021602(R) (2016).

\bibitem{Bisset PRA 16} R. N. Bisset, R. M. Wilson, D. Baillie, P. B. Blakie,
Phys. Rev. A \textbf{94}, 033619 (2016).


\bibitem{Wachtler PRA 16 a1} F. Wachtler and L. Santos, Phys. Rev. A \textbf{93}, 061603R (2016).


\bibitem{Wachtler PRA 16 a2} F. Wachtler and L. Santos, Phys. Rev. A \textbf{94}, 043618 (2016).


\bibitem{Blakie pra 16} P. B. Blakie, Phys. Rev. A \textbf{93}, 033644 (2016).





\bibitem{Boudjemaa PRA 20} A. Boudjemaa and N. Guebli, Phys. Rev. A \textbf{102}, 023302 (2020).


\bibitem{Heinonen PRA 19} V. Heinonen, K. J. Burns, and J. Dunkel, Phys. Rev. A \textbf{99}, 063621 (2019).


\bibitem{Malomed Phys D 19} B. A. Malomed, Physica D \textbf{399}, 108 (2019).


\bibitem{Shamriz PRA 20} E. Shamriz, Z. Chen, and B. A. Malomed, Phys. Rev. A \textbf{101}, 063628 (2020).


\bibitem{Li PRA 19} Z. Li, J.-S. Pan, and W. Vincent Liu, Phys. Rev. A \textbf{100}, 053620 (2019).


\bibitem{Aybar PRA 19} E. Aybar and M. O. Oktel, Phys. Rev. A \textbf{99}, 013620 (2019).


\bibitem{Examilioti JP B 20} P. Examilioti, and G. M. Kavoulakis, J. Phys. B: At. Mol. Opt. Phys. \textbf{53}, 175301 (2020).


\bibitem{Miyakawa PRA 20} T. Miyakawa, S. Nakamura, H. Yabu, Phys. Rev. A \textbf{101}, 033613 (2020).


\bibitem{Bottcher arXiv 20 07} F. Bottcher, Jan-Niklas Schmidt, J. Hertkorn,
Kevin S. H. Ng, Sean D. Graham, M. Guo, T. Langen, and T. Pfau, arXiv:2007.06391.

\bibitem{Bisset arXiv 20 07} R. N. Bisset, L. A. Peña Ardila, and L. Santos, arXiv:2007.00404.


\bibitem{Wang arXiv 20 02} Y. Wang, L. Guo, S. Yi, and T. Shi, arXiv:2002.11298.


\bibitem{Edmonds arXiv 20 02} M. J. Edmonds, T. Bland, and N. G. Parker, arXiv:2002.07958.


\bibitem{Baillie PRA 20} D. Baillie and P. B. Blakie, Phys. Rev. A \textbf{101}, 043606 (2020).



\bibitem{Dalfovo RMP 99} F. Dalfovo, S. Giorgini, L. P. Pitaevskii, and S. Stringari, Rev. Mod. Phys. \textbf{71}, 463 (1999).



\bibitem{Lee PR 57} T. D. Lee, K. Huang, and C. N. Yang, Phys. Rev. \textbf{106}, 1135 (1957).


\bibitem{Pitaevskii PRL 98} L. Pitaevskii and S. Stringari, Phys. Rev. Lett. \textbf{81}, 4541 (1998).

\bibitem{Braaten PRL 99} E. Braaten and J. Pearson, Phys. Rev. Lett. \textbf{82}, 255 (1999).

\bibitem{Astrakharchik PRL 05} G. E. Astrakharchik, R. Combescot, X. Leyronas, and S. Stringari, Phys. Rev. Lett. \textbf{95}, 030404 (2005).

\bibitem{Xu PRL 06} K. Xu, Y. Liu, D. E. Miller, J. K. Chin, W. Setiawan, and W. Ketterle, Phys. Rev. Lett. \textbf{96}, 180405 (2006).

\bibitem{Altmeyer PRL 07} A. Altmeyer, S. Riedl, C. Kohstall, M. J. Wright,
R. Geursen, M. Bartenstein, C. Chin, J. Hecker Denschlag, and R. Grimm,
Phys. Rev. Lett. \textbf{98}, 040401 (2007).

\bibitem{Papp PRL 08} S. B. Papp, J. M. Pino, R. J. Wild, S. Ronen, C. E. Wieman, D. S. Jin, and E. A. Cornell,
Phys. Rev. Lett. \textbf{101}, 135301 (2008).






\bibitem{Goral PRA 00} K. Goral, K. Rzazewski, and T.
Pfau, Phys. Rev. A \textbf{61}, 051601(R) (2000).

\bibitem{Santos PRL 00} L. Santos, G.V. Shlyapnikov, P. Zoller, and M.
Lewenstein, Phys. Rev. Lett. \textbf{85}, 1791 (2000).

\bibitem{Yi PRA 00} S. Yi, L. You, Phys. Rev. A, \textbf{61}, 041604(R) (2000).

\bibitem{Griesmaier PRL 05} A. Griesmaier, J. Werner, S. Hensler, J. Stuhler, and T. Pfau,
Phys. Rev. Lett. \textbf{94}, 160401 (2005).

\bibitem{Wang Nat 10} K.-K. Ni, S. Ospelkaus, D. Wang, G. Quemener, B. Neyenhuis,
M. H. G. de Miranda, J. L. Bohn, J. Ye, D. S. Jin, Nature (London) \textbf{464}, 1324 (2010).

\bibitem{Carr Ye NJP 09} L. D. Carr, J. Ye, New J. Phys. \textbf{11}, 055009 (2009).

\bibitem{Carr DeMille NJP 09} L. D. Carr, D. DeMille, R. V. Krems and J. Ye,
New J. Phys. \textbf{11}, 055049 (2009).


\bibitem{Wilson arxiv 11} R. M. Wilson, S. T. Rittenhouse and J. L. Bohn, New J. Phys. \textbf{14}, 043018 (2012).


\bibitem{Lahaye Nat 07} T. Lahaye, T. Koch, B. Frohlich, M. Fattori, J. Metz, A. Griesmaier, S. Giovanazzi, T. Pfau, Nature \textbf{448}, 672 (2007).

\bibitem{Lahaye RPP 09} T. Lahaye, C. Menotti, L. Santos, M. Lewenstein, and T. Pfau, Rep. Prog. Phys. \textbf{72}, 126401 (2009).

\bibitem{Lima PRA 10 a} A. R. P. Lima, A. Pelster, Phys. Rev. A \textbf{81}, 021606(R) (2010).
\bibitem{Lima PRA 10 b} A. R. P. Lima, A. Pelster, Phys. Rev. A \textbf{81}, 063629 (2010).





\bibitem{Lima PRA 11} A. R. P. Lima, A. Pelster, Phys. Rev. A \textbf{84}, 041604 (2011).

\bibitem{Lima PRA 12} A. R. P. Lima, A. Pelster, Phys. Rev. A \textbf{86}, 063609 (2012).

\bibitem{Blakie PRA 13} P. B. Blakie, D. Baillie, and R. N. Bisset, Phys. Rev. A \textbf{88}, 013638 (2013).



\bibitem{Andreev EPJ D 14} P. A. Andreev, L. S. Kuz'menkov, Eur. Phys. J. D \textbf{68}, 270 (2014).




\bibitem{Andreev 2005} P. A. Andreev, arXiv:2005.13503.

\bibitem{Andreev 2007} P. A. Andreev, arXiv:2007.15045.


\bibitem{Katsimiga NJP 17 01} G. C. Katsimiga, G. M. Koutentakis, S. I. Mistakidis, P. G. Kevrekidis,
and P. Schmelcher, New J. Phys. \textbf{19}, 073004 (2017).

\bibitem{Katsimiga NJP 17 02} G. C. Katsimiga, S. I. Mistakidis, G. M. Koutentakis, P. G. Kevrekidis
and P. Schmelcher, New J. Phys. \textbf{19}, 123012 (2017).

\bibitem{Katsimiga PRA 18} G. C. Katsimiga, S. I. Mistakidis, G. M. Koutentakis, P. G. Kevrekidis,
and P. Schmelcher, Phys. Rev. A \textbf{98}, 013632 (2018).

\bibitem{Mistakidis NJP 18} S. I. Mistakidis, G. C. Katsimiga, P. G. Kevrekidis, and P. Schmelcher, New J. Phys. \textbf{20}, 043052 (2018).




\bibitem{Koide PRC 13} T. Koide, Phys. Rev. C \textbf{87}, 034902 (2013).

\bibitem{Andreev EPL 16} P. A. Andreev, L. S. Kuz'menkov, Eur. Phys. Lett. \textbf{113}, 17001 (2016).

\bibitem{Andreev APL 16} P. A. Andreev, L. S. Kuz'menkov, Appl. Phys. Lett. \textbf{108}, 191605 (2016).



\bibitem{Andreev 2001} P. A. Andreev, arXiv:2001.02764.



\bibitem{Maksimov QHM 99} L. S. Kuz'menkov, S. G. Maksimov, Theor. Math. Phys. \textbf{118}, 227 (1999).
\bibitem{Andreev PRB 11} P. A.  Andreev, L. S. Kuzmenkov and M. I. Trukhanova, Phys. Rev. B \textbf{84}, 245401 (2011).
\bibitem{MaksimovTMP 2001} L. S. Kuz'menkov, S. G. Maksimov, and V. V. Fedoseev, Theor. Math. Phys. \textbf{126}, 110 (2001).
\bibitem{Renziehausen PTEP} K. Renziehausen, and I. Barth, Prog. Theor. Exp. Phys. \textbf{2018}, 013A05 (2018). 



\bibitem{Andreev 1912} P. A. Andreev, arXiv:1912.00843.

\bibitem{Andreev PRA08} P. A. Andreev, L. S. Kuz'menkov, Phys. Rev. A \textbf{78}, 053624 (2008).

\bibitem{Andreev LP 19} P. A. Andreev, Laser Phys. \textbf{29}, 035502 (2019).





\bibitem{Tokatly PRB 99} I. Tokatly, O. Pankratov, Phys. Rev. B \textbf{60}, 15550 (1999).


\bibitem{Tokatly PRB 00} I. V. Tokatly, O. Pankratov, Phys. Rev. B \textbf{62}, 2759 (2000).



\bibitem{Andreev 2003} P. A. Andreev, K. V. Antipin, M. Iv. Trukhanova, arXiv:2003.12547.



\bibitem{Bertini PRL 16} B. Bertini, M. Collura, J. De Nardis, and M. Fagotti, Phys. Rev. Lett. \textbf{117}, 207201 (2016).



\bibitem{Ruggiero QGH arxiv 19} P. Ruggiero, P. Calabrese, B. Doyon, and J. Dubail, Phys. Rev. Lett. \textbf{124}, 140603 (2020).




\bibitem{Sagdeev RPP 66} R. Z. Sagdeev, in Reviews of Plasma Physics, edited by M. A.
Leontovich (Consultants Bureau, New York, 1966), Vol. 4, p. 23.

\bibitem{Schamel PF 77} H. Schamel, M. Y. Yu, and P. K. Shukla, Phys. Fluids \textbf{20}, 1286 (1977).

\bibitem{Witt PF 83} E. Witt and W. Lotko, Phys. Fluids \textbf{26}, 2176 (1983).

\bibitem{Mamun PRE 97} A. A. Mamun, Phys. Rev. E \textbf{55}, 1852 (1997).

\bibitem{Shah PoP 10} H. A. Shah, M. N. S. Qureshi, and N. Tsintsadze, Phys. Plasmas \textbf{17}, 032312 (2010).

\bibitem{Marklund PRE 07} M. Marklund, B. Eliasson, and P. K. Shukla, Phys. Rev. E \textbf{76}, 067401 (2007).


\bibitem{Akbari-Moghanjoughi PP 17} M. Akbari-Moghanjoughi, Phys. Plasmas \textbf{24}, 052302 (2017).



















\end{thebibliography}
\end{document}